\documentclass[conference]{IEEEtran}
\IEEEoverridecommandlockouts
\usepackage{tikz}
\usepackage{xcolor}
\newcommand*\circled[1]{\tikz[baseline=(char.base)]{
\node[shape=circle,fill,inner sep=1pt] (char) {\textcolor{white}{#1}};}}
\usepackage{cite}
\usepackage{enumitem} 
\usepackage{amsmath,amssymb,amsfonts}
\usepackage{algorithmic}
\usepackage{graphicx}
\usepackage{textcomp}
\usepackage{xcolor}
\def\BibTeX{{\rm B\kern-.05em{\sc i\kern-.025em b}\kern-.08em
    T\kern-.1667em\lower.7ex\hbox{E}\kern-.125emX}}

\usepackage{soul}

\usepackage{subfig}
\usepackage{graphicx}
\usepackage{amssymb}    
\usepackage{ragged2e}
\usepackage{blindtext, graphicx}
\graphicspath{{Pictures/}{Profile/}} 

\begin{document}


\title{Enhancing Reliability of STT-MRAM Caches by Eliminating Read Disturbance Accumulation\\
}

\author{\IEEEauthorblockN{1\textsuperscript{st} Elham Cheshmikhani}
\IEEEauthorblockA{\textit{dept. of Computer Engineering} \\
\textit{Sharif University of Technology}\\
Tehran, Iran \\
elham.cheshmikhani@sharif.edu}
\and
\IEEEauthorblockN{2\textsuperscript{nd} Hamed Farbeh}
\IEEEauthorblockA{\textit{dept. of Computer Engineering and Information Technology } \\
\textit{Amirkabir University of Technology}\\
Tehran, Iran \\
farbeh@aut.ac.ir}
\and
\IEEEauthorblockN{3\textsuperscript{rd} Hossein Asadi}
\IEEEauthorblockA{\textit{dept. of Computer Engineering} \\
\textit{Sharif University of Technology}\\
Tehran, Iran \\
asadi@sharif.edu}
}

\maketitle

\begin{abstract}

$Spin$-\textit{$Transfer~Torque~Magnetic~RAM$} (STT-MRAM) as one of the most promising replacements for SRAMs in on-chip cache memories benefits from higher density and scalability, near-zero leakage power, and non-volatility, but its reliability is threatened by high read disturbance error rate.
$Error$-$Correcting~Codes$ (ECCs) are conventionally suggested to overcome the read disturbance errors in STT-MRAM caches.
By employing aggressive ECCs and checking out a cache block on every read access, a high level of cache reliability is achieved.
However, to minimize the cache access time in modern processors, all blocks in the target cache set are simultaneously read in parallel for tags comparison operation and only the requested block is sent out, if any, after checking its ECC.
These extra cache block reads without checking their ECCs until requesting the blocks by the processor cause the accumulation of read disturbance error, which significantly degrade the cache reliability.
In this paper, we first introduce and formulate the read disturbance accumulation phenomenon and reveal that this accumulation due to conventional parallel accesses of cache blocks significantly increases the cache error rate.
Then, we propose a simple yet effective scheme, so-called $Read~Error~Accumulation~Preventer~cache$ (REAP-cache), to completely eliminate the accumulation of read disturbances without compromising the cache performance.
Our evaluations show that the proposed REAP-cache extends the cache $Mean~Time~To~Failure$ (MTTF) by 171x, while increases the cache area by less than 1\% and energy consumption by only 2.7\%.
\end{abstract}

\begin{IEEEkeywords}
cache, STT-MRAM, read disturbance, Error-Correcting Code (ECC), error rate
\end{IEEEkeywords}

\section{Introduction}
$Static~Random~Access~Memories$ (SRAMs) have been the prevalent memory technology in on-chip caches.
SRAMs, however face several challenges, e.g, high leakage power and cell instability, by technology downscaling~\cite{tarihi2016hybrid}. 
Recent development in Non-Volatile Memory (NVM) technology has made
\textit{$Spin$}-\textit{$Transfer~Torque~Magnetic~RAM$} (STT-MRAM) as the most promising alternative for SRAMs in on-chip caches.
Near-zero leakage power, immunity to radiation-induced errors, higher density, better scalability, and non-volatility are the main advantages of STT-MRAM caches~\cite{EDCC}.
Beside the mentioned advantages, STT-MRAM caches are highly error-prone in read operations.
When a read current is applied to cache cells during a read operation, it is probable that the content of the cells unintentionally flips.
This error, known as \textit{read disturbance}, is originated from the stochastic switching behavior of STT-MRAM cells~\cite{Na-TCAS-II-16}.
Read disturbance error is a main source of unreliability in STT-MRAM caches.

Cache blocks are conventionally protected using $Error$-$Correcting~Codes$ (ECCs)~\cite{farbeh2016floating}.
On a request to a cache block, the ECC decoder logic checks the block for a possible correctable error and sends out the corrected data.
When the number of erroneous bits in a block is larger than the ECC correction capability, however, the ECC decoder fails to correct the data block.
The read operation of the caches in modern processors is optimized for access time reduction~\cite{farbeh2014psp, dai2014exploiting, sembrant2013tlc}, which significantly increases the occurrence probability of uncorrectable errors caused by read disturbances.

On a read request to a k-way set associative cache, in parallel to tag comparison operation and prior to determining the requested data block, all blocks in the target set is read simultaneously~\cite{mishra2015energy, mittal2014survey, sembrant2013tlc} and the requested block is sent out after checking its ECC.
The other $k-1$ blocks are discarded in this case.
These extra $k-1$ out of $k$ speculative reads from blocks that are not being requested in each read access increases the read disturbance probability in these blocks.
We call these extra imposed reads $concealed~reads$.
Reading multiple cache blocks for several times without checking their ECC causes the accumulation of read disturbance errors and occurrence of uncorrectable errors.
To the best of our knowledge, this significant reliability issue has not been addressed in the previous studies.

In this paper, 1) we first examine and formulate the adverse effect of concealed reads on the STT-MRAM cache reliability and analytically illustrate that the read disturbance accumulation severely increases the cache error rate. 
Then, 2) by conducting comprehensive experiments we show that the reliability of STT-MRAM cache is significantly degraded due to read disturbance accumulation.
Next 3), we propose an effective scheme, so-called $Read~Error~Accumulation~Preventer~cache$ (REAP-cache), to prevent the accumulation of read disturbance in cache blocks and completely eliminate the adverse effect of concealed reads on the cache reliability. 
REAP-cache scheme makes a minor modification in the cache read path to guarantee performing ECC checking not only for the requested block, but also for all other blocks within the cache set that have been read concurrently.

We evaluate REAP-cache using gem5 full-system simulator~\cite{gem5} running a set of workloads from SPEC CPU2006 benchmark suite~\cite{spec2006}. 
The results show that REAP-cache increases the $Mean~Time~To~Failure$ (MTTF) of STT-MRAM caches by 171x, on average.
The proposed scheme has no impact on the cache performance, increases the cache area by less than 1\%, and its energy consumption overhead is $only$ 2.7\%.


The rest of this paper is organized as follows. Section II describes the basics of STT-MRAM and its reliability challenges. In Section III, the observations and motivations for this work are discussed. The details of the proposed REAP-cache are presented in Section IV. Section V gives the simulation setup and evaluation results. Finally, we conclude the paper in Section VI.   

\section{STT-MRAM Basics}
\begin{figure}[b]
				\centering\vspace{-25pt}
				\subfloat[]{\includegraphics[width=0.43\linewidth]{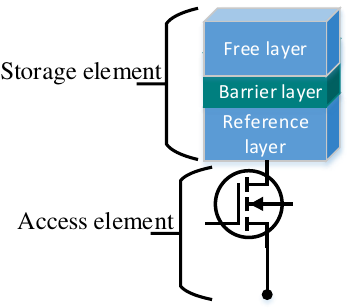}}
				\hspace{15pt}
				\subfloat[]{\includegraphics[width=0.27\linewidth]{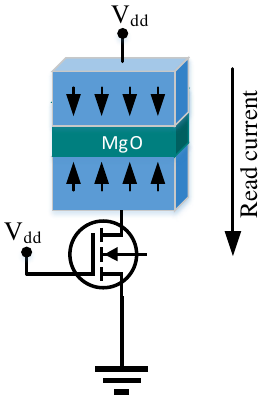}}\\
				\caption{(a) STT-MRAM cell structure with a MTJ unit and an access transistor and (b) Reading from STT-MRAM cell by applying read current.}
				\label{fig:1}
		\end{figure}

STT-MRAM cell consists of an access element and a storage element, as shown in Fig. \ref{fig:1}(a). 
The access element is an NMOS transistor used to connect and disconnect the cell to the array.
The other element is a magnetic storage element, named \textit{$Magnetic~Tunnel~Junction$} (MTJ), that stores data.
MTJ consists of three layers, a thin oxide barrier layer, which separates two ferromagnetic layers.
The oxide barrier layer, which is made up of crystallized MgO, is sandwiched between \textit{$reference~layer$} with a fixed magnetic field direction and \textit{$free~layer$} that its field direction can be changed through an applied current~\cite{EDCC}.
The magnetization direction of the free layer can be in the same or opposite direction as the reference layer.
If the spin-polarized current flow changes the orientation of the free layer spin to be parallel/anti-parallelism with reference layer spin, a low/high resistance is formed that is interpreted as storing logic value `0'/`1' in the cell.

Writing/reading to/from a STT-MRAM cell is done by applying a write/read current (\textit{I$_{write}$}/\textit{I$_{read}$}) for a predetermined pulse width.
To write `0' in the STT-MRAM cell, \textit{I$_{write}$} flows from bit line to source line to turn the electrons spin in the free layer in the same direction as that of in the reference layer.
Writing `1' in the cell needs a current flow from source line to bit line and causes the electron charge to flow from the free layer to the reference layer.
Reference layer with strong magnetic field reflects the electrons with the opposite direction to the free layer and these reflected electrons antiparallelize the magnetic field direction of the free layer and leads to write `1'.

To read a data from a cell, \textit{I$_{read}$} should flow either from bit line to source line or from source line to bit line to measure the MTJ resistance using a sense amplifier.
It means that, read operation is a unidirectional operation and is in the same direction as writing either `1' or `0'.
Fig. \ref{fig:1}(b) depicts the read operation from a STT-MRAM cell when the read current is adjusted to flow in the same direction as writing `0'.
In this way, during a read operation, the content of the cell is probable to unintentionally switch from `1' to `0' while \textit{I$_{read}$} flows through the cell.
This unexpected bit-flip that changes the cell content erroneously is named $read~disturbance$.
The probability of read disturbance occurrence in a STT-MRAM cell is calculated by~(\ref{eq:1})~\cite{Pajouhi2016JETC, AMir2016TDMR}.
\begin{flalign}
	\begin{split}
			\label{eq:1}
			 P_{Read-Disturbance} = 1- exp(\frac{-t_{read}}{\tau}\times \\
			{exp(\frac{-\Delta(I_{read}-I_{C_0})}{I_{C_0}}))}
			\end{split}
	\end{flalign}
Where, \textit{$\tau$} is attempt period and assumed to be 1ns, \textit{I$_{read}$} is read current, \textit{I$_{C0}$} is the current needed to write in 0$^{\circ}$K, \textit{t$_{read}$} is the read pulse width, and $\Delta$ is thermal stability factor.

There are some research studies that try to mitigate the read disturbance for improving STT-MRAM cells reliability.
The common method in architecture-level studies is $disruptive~reading~and~restoring$~\cite{wang2015selective,takemura2010highly}.
In these studies, a restore operation corrects the faulty bits after each read operation.
Therefore, disruptive~reading~and~restoring method increases both the cache access time and the $write~failure$ probability as another reliability challenge.
One of the studies that is categorized as a circuit-level method, try to leverage the read disturbance rate by minimizing the access transistor size~\cite{zhao2011design}. 
This resizing decreases the read and write current and decrease the read disturbance rate, while sensing this low current causes another reliability challenge known as $false~read$.
In \cite{na2016read,zhao2009high} the sense amplifiers are redesigned to tackle the inability of accurate sensing, while they increase the cell area and complicate the memory design.

\section{Observation and Motivation}
In this section, we demonstrate that the conventional cache configuration degrades the cache reliability by accumulating read disturbance occurrence in cache blocks in read accesses.
Then, we formulate the effect of read disturbance accumulation on the correctness of data blocks.
Finally, we evaluate the behavior of this accumulation in cache blocks to illustrate the severity of this problem and necessity for mitigating it.

\subsection{Read Disturbance Accumulation}
Considering a ECC-protected set-associative cache architecture, a sequence of operations is performed to commit a read request.
First, the cache set is determined using the requested address and the $tag$ part of the address is compared with tags of all blocks in the set to find a match.
On a cache hit, the requested block is read and is sent out after checking its ECC and on a cache miss, the procedure to handle the misses is conducted.
To minimize the cache access time, the logically sequential operations of tag comparison and reading from data block are conventionally performed in parallel.

In this configuration, known as $cache~parallel~access$ of $cache~fast~access$ mode~\cite{dai2013energy, park2012multistep, dai2014exploiting}, all data blocks have been read and ready for being sent out by the completion of the tag comparison operation.
On a cache hit, the output of the tag comparison selects between the available data blocks and the other data blocks are discarded.
This parallel cache access eliminates the time of accessing and reading from data block in the cost of higher energy consumption due to extra reads from data blocks.
For each access, $k$ data blocks are read in response to a request for a single block.

\begin{figure}[t]
				\centering\vspace{-14pt}
				\includegraphics[width=0.88\linewidth]{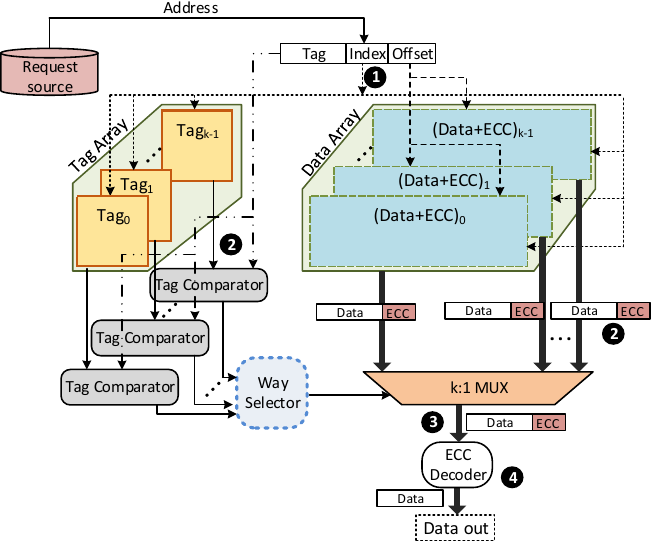}
				\caption{Cache structure in conventional parallel access mode.}\vspace{-15pt}
				\label{fig:2}
	      \end{figure}

Fig. \ref{fig:2} illustrates a conventional k-way set-associative cache  and the sequence of operations performed in cache fast access mode.
In step~\circled{1}, the index part of the input address is extracted to determine the target cache set.
In step~\circled{2}, all tags in the selected set is read and compared with the tag part of the input address.
At the same time, all cache lines in the selected set are speculatively read and delivered to the inputs of the MUX unit whose control signals are the output of the tag comparison unit.
In step~\circled{3}, on a cache hit, the MUX unit delivers one out of $k$ data blocks to the ECC decoder unit and the other $k-1$ data blocks are discarded, ehereas on a cache miss all $k$ data blocks are discarded.
The selected data block on a cache hit is checked by ECC decoder unit for a possible error and data is sent out at step~\circled{4}.

As described, all $k$ cache lines in a set are read in response to a request for reading a single line, while only the data of requested line is checked for possible error.
The other lines are read and under read disturbance error without being checked by ECC decoder.
These extra reads, named $concealed~read$ in this study, accumulate the read disturbance errors in the lines until the line is being requested.

A cache line can be under several hundreds or even thousands number of concealed reads and accumulation of read disturbance errors between two conservative results and checking its ECC.
If the number of erroneous bits due to accumulation of read disturbance errors in a cache line exceeds the error correction capability of ECC, the data cannot be recovered.
The next section formulates the occurrence probability of uncorrectable errors in cache lines.

\subsection{Problem Formulation}

An ECC-protected cache is conventionally capable of correcting single bit error in cache lines.
As mentioned, unidirectional read disturbance only affects STT-MRAM cells containing logic value `1'.
The probability of correct data block delivery when reading from a line is according to (\ref{eq:2}).
 \begin{flalign}
	\begin{split}
			\label{eq:2}
			 P_{corr-blk} = {(1-P_{RD-cell})}^n \\
			 +n\times P_{RD-cell}\times {(1-P_{RD-cell})}^{n-1} 
			\end{split}
	\end{flalign}
Where, $P_{RD-cell}$ is the probability of read disturbance occurrence in a cell on a read access (given in (1)) and $n$ is the number of `1' in the cache line.
Equation (\ref{eq:2}) is a binomial experiment in which $P_{RD-cell}$ is the probability of failure in each Bernoulli trial, $n$ is the number of trials and the probability of correct data block delivery ($P_{corr_{blk}}$) is the binomial probability of at most one failure.

      \begin{figure*}[t]
				\centering
				\subfloat[]{\includegraphics[width=0.45\linewidth]{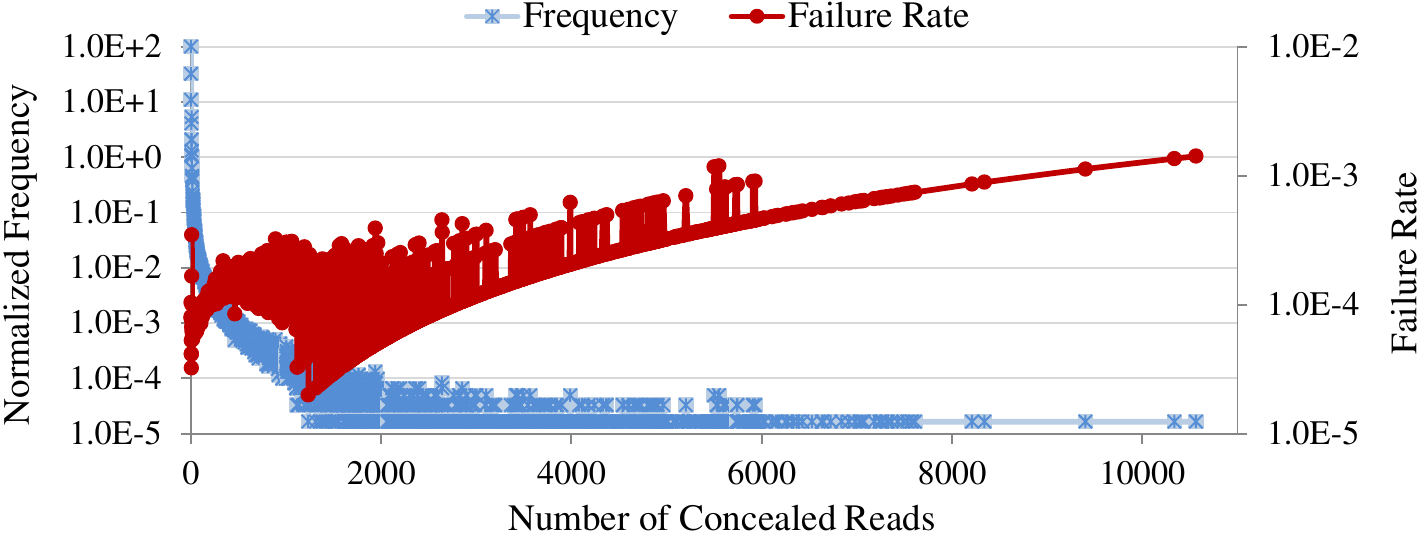}}
				\hspace{15pt}
				\subfloat[]{\includegraphics[width=0.45\linewidth]{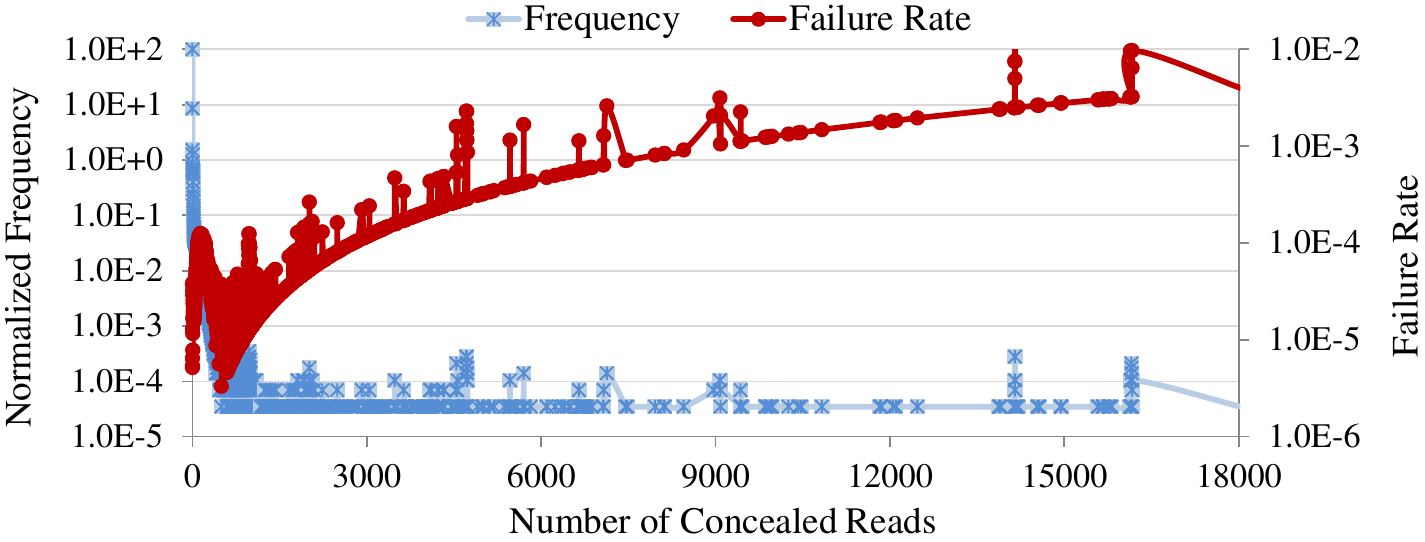}}\\
				\subfloat[]{\includegraphics[width=0.45\linewidth]{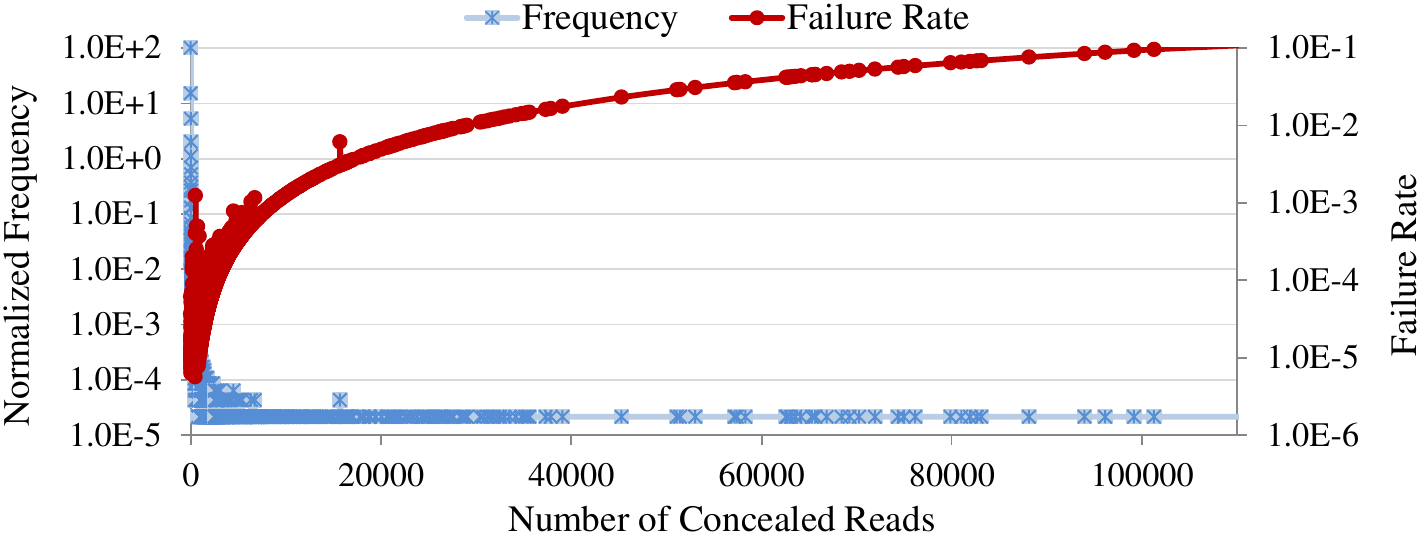}}
				\hspace{15pt}
				\subfloat[]{\includegraphics[width=0.45\linewidth]{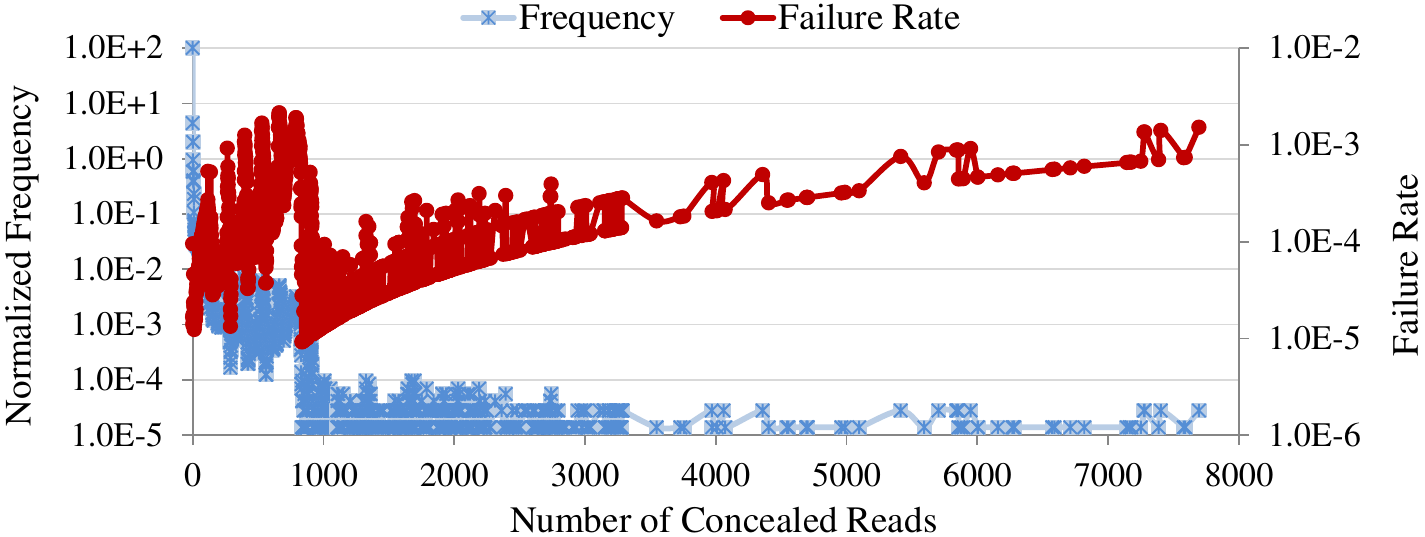}}
				\caption{Frequency of each number of concealed reads for cache access during the workload execution and the contribution of each frequency in total cache failure rate (a) perlbench, (b) calculix, (c) h264ref, and (d) dealII workload.}\vspace{-15pt}
				\label{fig:3}
			\end{figure*}
			
Equation (\ref{eq:2}) depicts that the correct data is delivered if all STT-MRAM cells in the line are error-free or only one out of $n$ cells is erroneous.
In other words, ECC decoder unit is capable of delivering the correct data $if~and~only~if$ at most one data cell is erroneous.
As mentioned, several concealed reads are probable for a cache line before requesting it and checking its ECC.

The read disturbance probability in each concealed read is accumulated in the line and degrade the probability of correct data delivery, which is calculate according to~(\ref{eq:3}).
\begin{flalign}
	\begin{split}
			\label{eq:3}
			 P_{corr-blk-acc} = (1-P_{RD-cell})^{N\times n} \\
			 +N\times n\times P_{RD-cell}\times (1-P_{RD-cell})^{N\times n-1} 
			\end{split}
	\end{flalign}
Where, $N$ is the number of concealed reads occurred between two consecutive accesses to the line plus one (to count the last read access).
Interpreting as a binomial experiment, by comparing~(\ref{eq:2}) and~(\ref{eq:3}), it is evident that concealed reads increase the number of trials from $n$ to $N\times n$ while the number of failures still must be at most one in binomial probability.

We further clarify the effects of concealed reads on the occurrence probability of uncorrectable reads by a numerical example.
Assume that 100 bits out of 512 bits of a cache line contain logic value `1' and $P_{RD-cell}=10^{-7}$.
The probability of uncorrectable error in this line after a read access without any concealed read is according to~(\ref{eq:4}):
\begin{flalign}
	\begin{split}
			\label{eq:4}
			 P_{err} = 1-P_{corr-blk}=1-((1-10^{-8})^{100}+  \\
			  100\times 10^{-8}\times (1-10^{-8})^{99})=5.0\time 10^{-13}    
			\end{split}
	\end{flalign}
and probability of uncorrectable error considering 50 concealed reads is according to~(\ref{eq:5}): 
\begin{flalign}
	\begin{split}
			\label{eq:5}
			 P\prime_{err-acc} = 1-((1-10^{-8})^{100\times 50}+  \\
			  50\times 100\times 10^{-8}\times (1-10^{-8})^{100\times 50 -1}) =1.3\times 10^{-9}
			 \end{split}
	\end{flalign}
The above example illustrates that only 50 concealed read increases the probability of inability to deliver a correct data block by more than 3 orders of magnitude.

The above analysis is supported by a set of experiments for an 8-way set-associative L2 cache evaluated in gem5 full-system simulator~\cite{gem5}, using workloads from SPEC CPU2006 benchmark suite~\cite{spec2006}.
The evaluations show that the number of concealed reads in cache lines can be even higher than $10^5$ in some workloads, which extremely increases the occurrence probability of read disturbance in more than one bit.

Fig.~\ref{fig:3} shows the frequency of the number of concealed reads for all read accesses during the workload execution as well as their contribution in total cache failure probability, for four exemplary workloads.
More precisely, x-axis is the number of concealed reads and the primary y-axis is the number of read requests to all lines with each specific number of concealed reads, which is normalized to the number of read accesses from lines with no concealed read.
For example, point (35, 3) means that 3 lines are read with 35 concealed reads during the workload execution, if the number of reads from lines without any concealed read is scaled to 100.
The secondary y-axis depicts the failure probability considering both the number of concealed reads and frequency of each concealed reads.
This probability has direct relation with both frequency and the number of concealed reads.
Both primary and secondary y-axes are in logarithmic scale.

%

The results show that the frequency of reading from lines with a each concealed read decreases for higher number of concealed reads, while the contribution of lines with high number of concealed reads in total cache failure rate is significantly higher.
For example, in $perlbench$ workload depicted in Fig.~\ref{fig:3}(a), the frequency of reading from lines with 8000 concealed reads is by six orders of magnitude lower than that of lines with less than five concealed reads, while the cache failure probability due to the former is higher than that of the latter by two orders of magnitude.

The same trend is observed in $calculix$ depicted in Fig.~\ref{fig:3}(b), in which the cache failure rate due to line with $10^4$ concealed reads is by more than 100x higher than that of lines with less than 20 concealed reads, despite the extremely lower frequency of the former.
The number of concealed read in $h264ref$ workload exceeds $10^5$ for some cache lines, which makes them the main contributor in total cache failure rate, according to Fig.~\ref{fig:3}(c).
The cache lines with higher number of concealed reads shown in Fig. \ref{fig:3}(d) also are the dominant contributor in cache failure rate, for $dealII$ workload.

The above observations demonstrate that concealed reads in conventional cache configuration significantly degrade the STT-MRAM reliability in the sake of higher cache performance.
In the next section, we propose a scheme to eliminate the impact of concealed reads on the reliability without affecting the cache performance.

\section{Proposed Method}

In a conventional k-way set-associative cache structure, each read from a line in a set imposes $k-1$ extra reads from other lines in that set.
These concealed reads are discarded without checking their ECCs.
This causes read disturbance error accumulation and decreases the probability of correct read operation.
To confront this challenge and eliminate the adverse effect of concealed reads, two approaches can be imagined:
1) reading a data line after completion of tag comparison operation, while removes the performance benefit of cache parallel access and significantly increases the cache access time, 
2) performing ECC checking for all data lines on a read access instead of only checking the requested line.

Our proposed scheme is based on the second approach and guarantees to prevent read disturbance accumulation without affecting the cache performance.
Taking into account the cache operation on the read path described in Fig.~\ref{fig:2}, we propose the ultra low-cost $Read~Error~Accumulation~Preventer$-$cache$ (REAP-cache) scheme, which makes a minor modification in the cache structure.
This modification provides the opportunity of simultaneously checking the ECC of all data lines accessed during concealed a read in addition to the requested line.

\begin{figure}[t]
				\centering\vspace{-14pt}
				\includegraphics[width=0.9\linewidth]{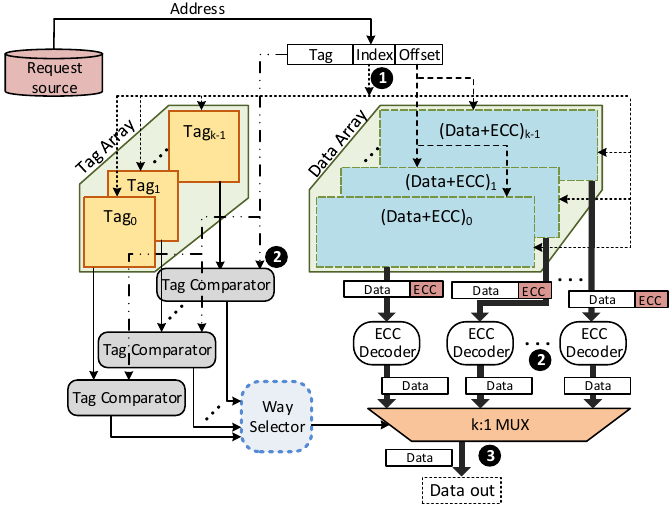}
				\caption{Cache structure in proposed REAP-cache.}\vspace{-15pt}
				\label{fig:4}
	      \end{figure}

Fig.~\ref{fig:4} depicts the cache structure in REAP-cache scheme.
Referring to the conventional cache structure in Fig.~\ref{fig:2}, in REAP-cache scheme, we swap the location of ECC decoder unit and the MUX unit in the read path.
To make it possible to simultaneously check all data blocks, we replicate the ECC decoder unit equal to number of lines accessed in parallel.
By this modification in the read path, we garantee to perform ECC checking for all cache line after each read operation regardless of the type of read, i.e., concealed or real read.
It can be theoretically proven that the cache access time is not increased by this swapping of MUX and ECC decoder unit.

The sequence of cache operations in a read request is depicted in Fig.~\ref{fig:4}.
Same as in the conventional caches, the target set is determined by index part of the input address in step~\circled{1}.
In step~\circled{2}, all tags in the set are read and compared with the input tag and all data lines are read simultaneously.
All $k$ data blocks are sent to $k$ ECC decoding units and the corrected data are delivered to MUX unit by the end of this step.
The output of tag comparison unit selects one of the data blocks in step~\circled{3} and the data is sent out by the cache.

REAP cache scheme eliminates the accumulation of read disturbance errors by transforming the checking a cache line after $N$ concealed reads into $N$ times checking the line each after a read operation.
In this case, the probability of uncorrectable error in the line will be according to~(\ref{eq:6}):
\begin{flalign}
	\begin{split}
			\label{eq:6}
			 P\prime_{corr-blk-REAP} = ((1-P_{RD-cell})^n  \\
			  +n\times P_{RD-cell}\times {(1-P_{RD-cell})}^{n-1} )^N
			\end{split}
	\end{flalign}

Considering our previous example in Section III for a line with 100 bits of `1' and 50 concealed reads, the probability of uncorrectable error is $2.6\times 10^{-11}$, which is 50x lower than that of conventional cache.

\section{Simulation Setup and Results}
To evaluate the proposed REAP-cache scheme, we modeled a processor containing two levels of on-chip caches in gem5 full-system simulator~\cite{gem5}.
L1 cache is comprised of dedicated data and instruction cache with SRAM technology and L2 cache is shared with STT-MRAM technology.
The details of cache configuration is given in Table~\ref{table:1}.
We use SPEC CPU2006 benchmark suite~\cite{spec2006} as the workload and extract the result by running one billion instructions after skipping the initial 100 million warm-up instructions.
REAP-cache scheme is compared with the conventional cache as the baseline in terms of reliability, energy consumption, area, and performance.
As a reliability metric, we calculated the Mead Time to Failure (MTTF) of cache in REAP-cache and normalized it to MTTF in the baseline.
The STT-MRAM cache is modeled in NVSim simulator~\cite{nvsim} to extract the energy consumption, area, and access time parameters.
In the following, we first investigate the cache reliability and then explore the overheads of REAP-cache scheme.

\begin{table}[t]\vspace{-10pt}
				\centering
				\caption{Configuration of On-Chip Caches}\vspace{-10pt}
				\includegraphics[width=1\linewidth]{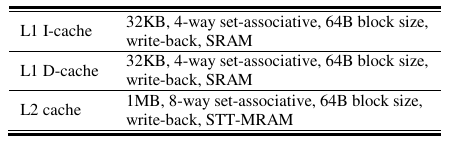}
				\label{table:1}\vspace*{-20pt}
			\end{table}
\begin{figure}[t]
				\centering\vspace{5pt}
				\includegraphics[width=0.91\linewidth]{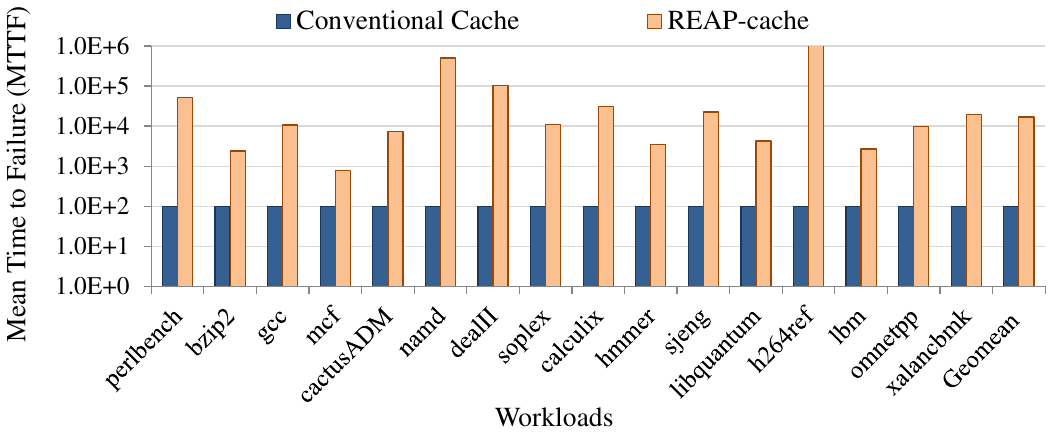}
				\caption{Mean Time To Failure of cache in REAP-cache normalized to conventional cache.}\vspace{-15pt}
				\label{fig:5}
	      \end{figure}

\subsection{Reliability Evaluation}
REAP-cache scheme increases the cache reliability by eliminating the accumulation of read disturbance errors.
Read disturbance error is probable for each read access in a cache line and when the line is not check for several accesses, these errors are accumulated and the ECC decoder is not capable of delivering correct data if the number of errors exceeds the ECC correction capability.
By performing ECC checking after each read access to a line in REAP-cache scheme, ECC decoder needs to only correct the errors occurred in a single read access, which results in a higher error correction capability and lower cache failure rate.
Fig.~\ref{fig:5} depicts the MTTF of the STT-MRAM cache in REAP-cache scheme normalized to the baseline for all workloads.
The results show that REAP-cache improves the MTTF of the cache by 171x, on average. This value for some workloads is more than 1000x, e.g., $namd$, $dealII$, $h264ref$. In the worst case, REAP-cache increases the MTTF by 7.9x in $mcf$ workload.
Increasing the MTTF of the cache is equivalent to reducing its error rate by an average of 171x, which is achieved by providing higher error correction capability in REAP-cache.

\subsection{Overhead}
REAP-cache has no effect on the cache energy consumption in write operations as well as in tag unit operations of the cache. However, REAP-cache increases the cache energy consumption by performing larger number of ECC checking per read access. There is single ECC decoder unit in conventional cache structure for checking the requested cache line, whereas REAP-cache requires eight ECC decoder units for simultaneously checking all data blocks in a cache set. The contribution of ECC decoder unit in total energy consumption of the cache is less than 1\% and the impact of the modification made by REAP-cache in cache structure on the energy consumption is not significant.

\begin{figure}[t]
				\centering\vspace{5pt}
				\includegraphics[width=0.91\linewidth]{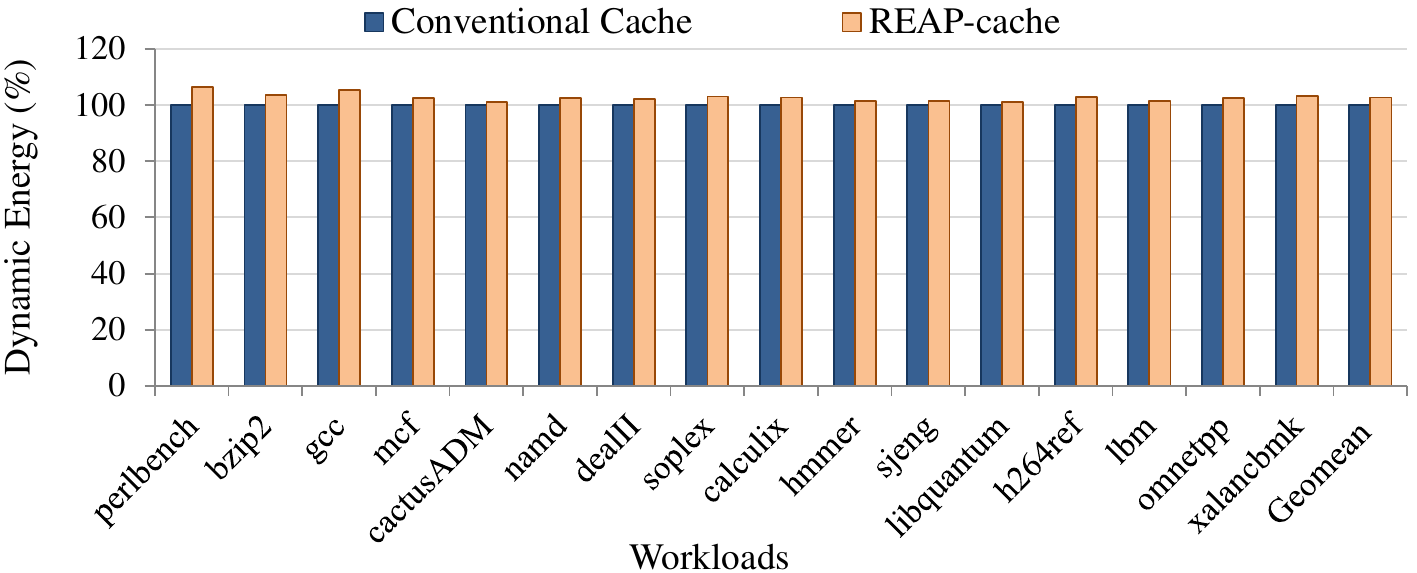}
				\caption{Dynamic energy consumption in REAP-cache normalized to conventional cache.}\vspace{-15pt}
				\label{fig:6}
	      \end{figure}

Fig.~\ref{fig:6} shows the dynamic energy consumption of STT-MRAM cache in REAP-cache normalized to the baseline for all workloads.
According the results, REAP-cache increases the energy consumption by an average of 2.7\%.
The worst case overhead of 6.5\% is observed in cactusADM workload and in the best case, REAP-cache imposes 1.0\% overhead in $xalancbmk$ workload.
The variations in the energy consumption overheads is mainly due to different contributions of read accesses in total energy consumption.
The fraction of read accesses in total cache accesses as well as fraction of dynamic energy in total energy consumption varies for different workloads, which leads to a small variation in the overhead of REAP-cache.

In term of area, the overhead of REAP-cache is only due to including ECC decoder units.
Based on our evaluations, the contribution of ECC decoder unit in total cache area is about 0.1\%.
Therefore, the area overhead due to increasing the number of ECC decoder units from one to eight in an 8-way set-associative cache is less than 1\%, which is negligible.

From performance point of view, REAP-cache has no adverse effect on the cache access time.
By comparing the cache structure in Fig.~\ref{fig:2} and Fig.~\ref{fig:4}, it is evident that read path access time in REAP-cache is less than or equal to that of the baseline.
By swapping the MUX unit and ECC decoder unit in the read path, REAP-cache provides the opportunity to overlap not only the data lines accesses, but also the ECC decoding operation with the tag comparison operation. Hence, the cache access time can be even reduced in REAP-cache.

\section{Conclusion}
Cache lines are accessed in parallel to tag comparison operation for minimizing the cache access time in modern processors. This paper revealed that the consequence of this optimization is a previously unknown reliability problem, i.e., read disturbance accumulation, when replacing SRAMs with STT-MRAM technology in on-chip caches. Parallel accesses to all cache lines before determining the requested line significantly degrade the reliability of the cache by increasing the occurrence probability of uncorrectable errors. We proposed REAP-cache scheme to completely eliminate the accumulation of read disturbance errors. By making a minor modification in conventional cache structure, REAP-cache extends the MTTF of the STT-MRAM caches by an average of 171x. This significant reliability improvement is achieved with no performance degradation, imposing less than 1\% area overhead, and increasing the cache energy consumption by only 2.7\%.

\def\bibfont{\footnotesize}\footnotesize       
\bibliographystyle{IEEEtran}
\bibliography{IEEEabrv,references}
%
%
%
%
\end{document}